\documentclass[fleqn,12pt]{wlscirep}
\usepackage{aas_macros}
\usepackage{setspace}
\usepackage{subcaption}
\usepackage{caption}

\title{The corona contracts in a new black hole transient}

\author[1,2,3,4,*]{E. Kara}
\author[4]{J. F. Steiner}
\author[5]{A. C. Fabian}
\author[6]{E. M. Cackett}
\author[7]{P. Uttley}
\author[4]{R. Remillard}
\author[2]{K. Gendreau}
\author[2]{Z. Arzoumanian}
\author[8]{D. Altamirano}
\author[9,10]{S. Eikenberry}
\author[11]{T. Enoto}
\author[12,13]{J. Homan}
\author[14]{J. Neilsen}
\author[15]{A. L. Stevens}

\affil[1]{University of Maryland, College Park, MD 20784, USA}
\affil[2]{NASA Goddard Space Flight Center, Greenbelt, MD}
\affil[3]{Joint Space Science Institute, University of Maryland, College Park, MD, 20742}
\affil[4]{MIT Kavli Institute for Astrophysics and Space Research, Cambridge, MA 02139, USA}
\affil[5]{Institute of Astronomy, Madingley Road, Cambridge CB3 0HA, UK}
\affil[6]{Wayne State University, Department of Physics \& Astronomy, Detroit, MI 48201, USA}
\affil[7]{Anton Pannekoek Institute for Astronomy, University of Amsterdam, Science Park 904, 1098 XH Amsterdam, NL}
\affil[8]{School of Physics and Astronomy, University of Southampton, Southampton, SO17 1BJ, UK}
\affil[9]{Department of Astronomy, University of Florida, Gainesville, FL 32611, USA}
\affil[10]{Department of Physics, University of Florida, Gainesville, FL 32611, USA}
\affil[11]{Hakubi Center for Advanced Research and Department of Astronomy, Kyoto University, Kyoto 606-8302, Japan}
\affil[12]{Eureka Scientific, Inc., 2452 Delmer Street, Oakland, CA 94602, USA}
\affil[13]{SRON, Netherlands Institute for Space Research, Sorbonnelaan 2, 3584 CA Utrecht, NL}
\affil[14]{Villanova University, Department of Physics, Villanova, PA 19085, USA}
\affil[15]{Department of Physics \& Astronomy, Michigan State University, 567 Wilson Road, East Lansing, MI 48824, USA}
\affil[*]{ekara@astro.umd.edu}

\vspace{-2cm}
\begin{abstract}

\textbf{The geometry of the accretion flow around stellar-mass black holes can change on timescales of days to months\cite{remillard06,fender04,done07}. When a black hole emerges from quiescence it has a very hard X-ray spectrum produced by a hot corona\cite{rees82, narayan96}, and then transitions to a soft spectrum dominated by emission from a geometrically thin accretion disc extending to the innermost stable circular orbit\cite{shakura73, steiner10}. Much debate, however, persists over how this transition occurs, whether it is driven largely by a reduction in the truncation radius of the disc\cite{plant14,ingram11} or in the spatial extent of the corona\cite{fabian14,garcia15}. Observations of X-ray reverberation lags in supermassive black hole systems\cite{fabian09,uttley14} suggest that the corona is compact and that the disc extends in close to the central black hole\cite{zoghbi12,kara16}. Observations of stellar mass black holes, however, reveal equivalent (mass-scaled) reverberation lags that are much larger\cite{uttley11}, leading to the suggestion that the accretion disc in the hard state of stellar mass black holes is truncated out to hundreds of gravitational radii\cite{demarco15,demarco17}. Here we report X-ray observations of the new black hole transient MAXI J1820+070\cite{kawamuro18,uttley18}. We find that the reverberation time lags between the continuum-emitting corona and the irradiated accretion disc are 6–20 times shorter than previously seen. The timescale of the reverberation lags shortens by an order of magnitude over a period of weeks, while the shape of the broadened iron K emission line remains remarkably constant. This suggests a reduction in the spatial extent of the corona, rather than a change in the inner edge of the accretion disc.}

\end{abstract}
\begin{document}

%\flushbottom
\maketitle
%\thispagestyle{empty}

%\vspace{-0.2cm}

\begin{figure}
\centering
\includegraphics[width=\linewidth]{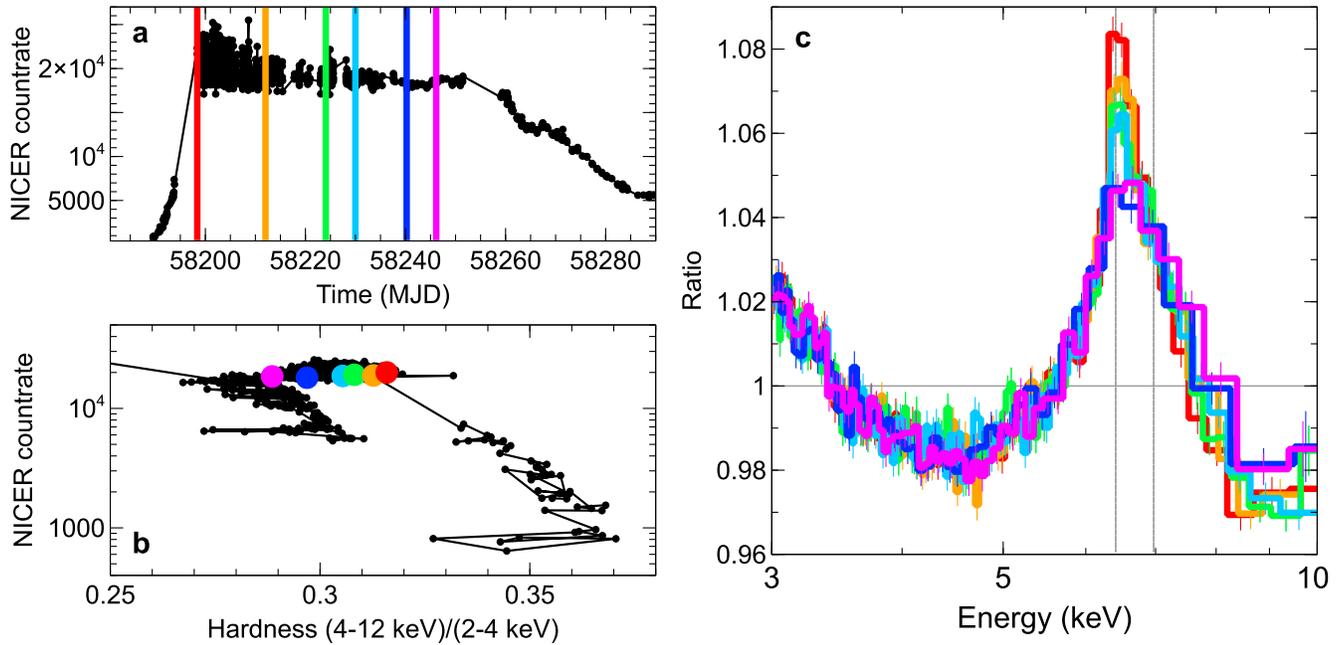}
\caption{\textbf{Overview of MAXI~J1820+070 in the hard state.} (a) The long term 0.2-12~keV NICER light curve of MAXI~J1820+070 in the hard state (black) overplotted with the times of the six {\em NICER} observations that are the subject of this analysis. After MJD~58290, the source began to rapidly transition to the soft state; \ref{fig:other_lags}-b. (b) The NICER Hardness-Intensity Diagram, defined as the total 0.2--12 keV count rate vs. the ratio of hard (4--12 keV) / soft (2--4 keV) count rates. The black points show MAXI~J1820+070 from its time of discovery to 2018 June 20. The six epochs of interest are shown as the colored points. We focus on a luminous phase, during which the source became gradually softer with time, but remained always in the hard, corona-dominated state. Error bars are similar to the size of the points. (c) The corresponding spectra of the six epochs fit to a simple powerlaw in the 3--10~keV range (with normalization and photon index free to vary between observations). The ratio plot reveals a clear broad iron~K emission line, and a narrow component at 6.4~keV that decreases in equivalent width as the source evolves in time. Thin gray vertical lines indicate Fe K$\alpha$ at 6.4~keV and H-like Fe XXVI at 6.97~keV. Error-bars indicate 1-$\sigma$ confidence intervals.}
\label{fig:HID_spec}
\end{figure}

MAXI~J1820+070\cite{kawamuro18} (ASASSN-18ey\cite{tucker18}) was discovered on 2018 March 11 with the Monitor of All-sky X-ray Image (MAXI) on board the International Space Station. The next day, the Neutron star Interior Composition Explorer (NICER)\cite{gendreau16} started obtaining detailed observations and has continued observing since, at a cadence of 1-3~days\cite{uttley18}. The NICER X-ray Timing Instrument consists of an aligned collection of 52 active paired X-ray ``concentrator'' optics and silicon drift detectors, which record the arrival times and energies of individual X-ray photons. It provides a timing resolution of $<100$ ns (25x faster than NASA's previous best X-ray timing instrument, the Rossi X-ray Timing Explorer) and the highest ever soft band peak effective area of 1900 cm$^{2}$ (nearly twice that of timing-capable EPIC-pn camera on XMM-Newton), all while providing good spectral resolution (145~eV at 6~keV), minimal pile-up on bright sources and very little deadtime.
MAXI~J1820+070 regularly reached 25000 counts/s in NICER's 0.2-12 keV band, while still providing high-fidelity spectral and timing products (for comparison, the XMM-Newton detectors become piled up for count rates of 600-800 counts/s\cite{diaztrigo12}). This high count rate allows us to probe timescales that are nearly an order of magnitude shorter than possible with XMM-Newton. Due to the enormity of the dataset, in this letter, we only describe the spectral-timing results of a subset of the total NICER observations (\ref{tab:obs} and Fig.~\ref{fig:HID_spec}a,b) when the source was brightest ($L_{\mathrm{0.3-10~keV}}/L_{\mathrm{Edd}} = 0.04-0.06$, using the parallax distance measure of $\sim3.3$~kpc\cite{homan18,gandhi18} and assuming a $10~M_{\odot}$ black hole). More detail on the remaining observations can be found in the Methods section.

Fig.~\ref{fig:HID_spec}c shows a simple ratio of the spectra from all six epochs to a powerlaw model fit to the 3--10~keV band. The photon index and normalization are left free to vary between observations. (see Methods for a description of the data reduction). All six epochs show a remarkably constant broad iron~K emission line that extends down below 5 keV. This is best fit by relativistic reflection from a point source X-ray corona at at a height of $< 5 r_{\mathrm{g}}$ irradiating a disc with an inner radius of $< 2 r_{\mathrm{g}}$. In addition, there is a narrow feature at 6.4~keV that is clearly present at early times (with an equivalent width of $\sim 50$~eV), but is less prominent in later epochs (down to an equivalent width of $\sim 10$~eV). The spectral modeling of MAXI~J1820+070 will be presented in detail in Fabian et al., in preparation. 

\begin{figure}
\centering
\includegraphics[width=\linewidth]{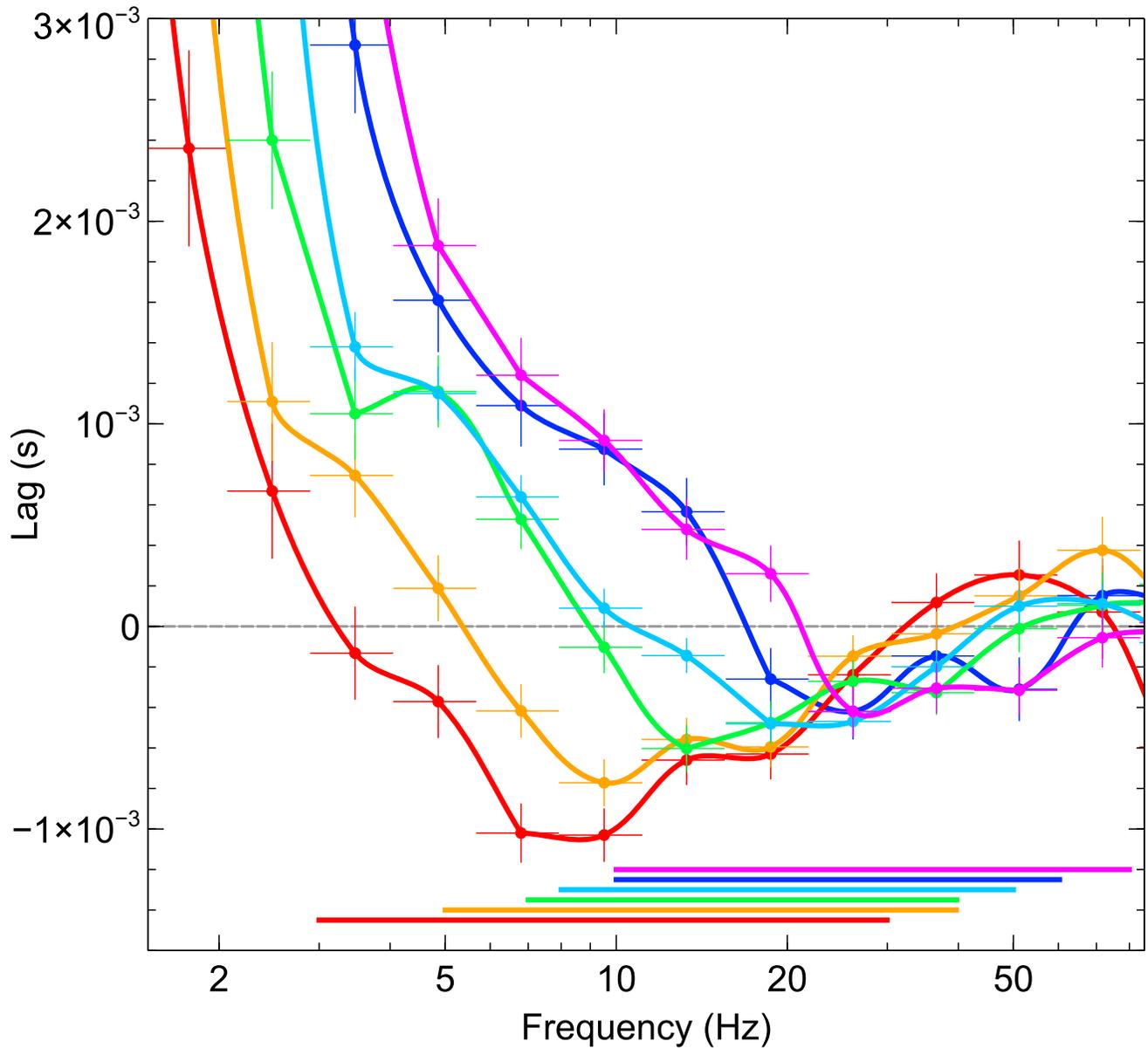}
\caption{\textbf{The evolution of the lag-frequency spectra.} The evolution of the lag between 0.5--1~keV and 1--10 keV as a function of temporal frequency for our six observation epochs. Color-coding is the same as Fig.~\ref{fig:HID_spec}. The points have been connected with a Bezier join to guide the eye. A negative lag indicates that the soft band follows behind the hard. The soft lag evolves to higher frequencies with time. The solid lines on the bottom portion of the figure indicate the frequencies used in the lag-energy analysis (Fig.~\ref{fig:lagen}). Error-bars indicate 1-$\sigma$ confidence intervals.}
\label{fig:lagfreq}
\end{figure}

In order to explore the time-dependence of these spectral signatures, we perform a frequency-resolved timing analysis (see Methods section for details and \ref{fig:psd} for the 0.01-100~Hz power spectrum of each epoch). We examine the frequency-dependent time lag between the 0.5--1~keV and 1--10~keV emission for the 6 epochs (Fig.~\ref{fig:lagfreq}). At low Fourier frequencies (at a few Hz and below), we observe a positive lag, defined as the hard band following after the soft, at all epochs (see Methods; \ref{fig:low_freq_lagen}). The low-frequency hard lag is a near ubiquitous feature of Galactic black hole binaries in the hard and intermediate states\cite{miyamoto89,altamirano15} and also of Type 1 AGN\cite{papadakis01}. The low-frequency hard lags are commonly interpreted as due to mass accretion rate fluctuations in the disc that propagate inwards on a viscous timescale causing the response of soft photons before hard\cite{kotov01}.

Th\textbf{}e lags at high frequencies show a reversal of sign, where the soft band begins to lag behind the hard. The soft lag is found in all observations at $4.5\sigma$ confidence or greater (see Methods for details). This suppression of the hard continuum lag is often seen in AGN systems, but is rarely seen in Galactic black hole binaries, where usually the hard lag continues to dominate over all variability timescales that can be probed. High-frequency soft lags are commonly interpreted as due to reflection off the inner accretion flow. In Galactic black hole binaries, the hard X-ray corona irradiates the accretion disc, reheating the disc and causing a lag of the thermal emission on the shortest timescales\cite{uttley11}. In these epochs (and confirmed in the other NICER observations; \ref{fig:other_lags}-a), we observe a trend of the soft lag towards progressively shorter timescales, suggesting an evolution in the accretion flow itself. This evolution towards a smaller emitting region has been inferred both spectroscopically\cite{plant14,garcia15}, and separately through timing properties\cite{ingram11,demarco15}, but there remains strong debate on the absolute size scale of the emitter, and whether it is the corona that is becoming more compact or the truncation radius of the disc that is decreasing. The evolution of the thermal reverberation lags to higher frequencies, together with the unchanging shape of the iron line profile, suggest that the evolution is driven by the corona.

\begin{figure}
\centering
\includegraphics[width=\linewidth]{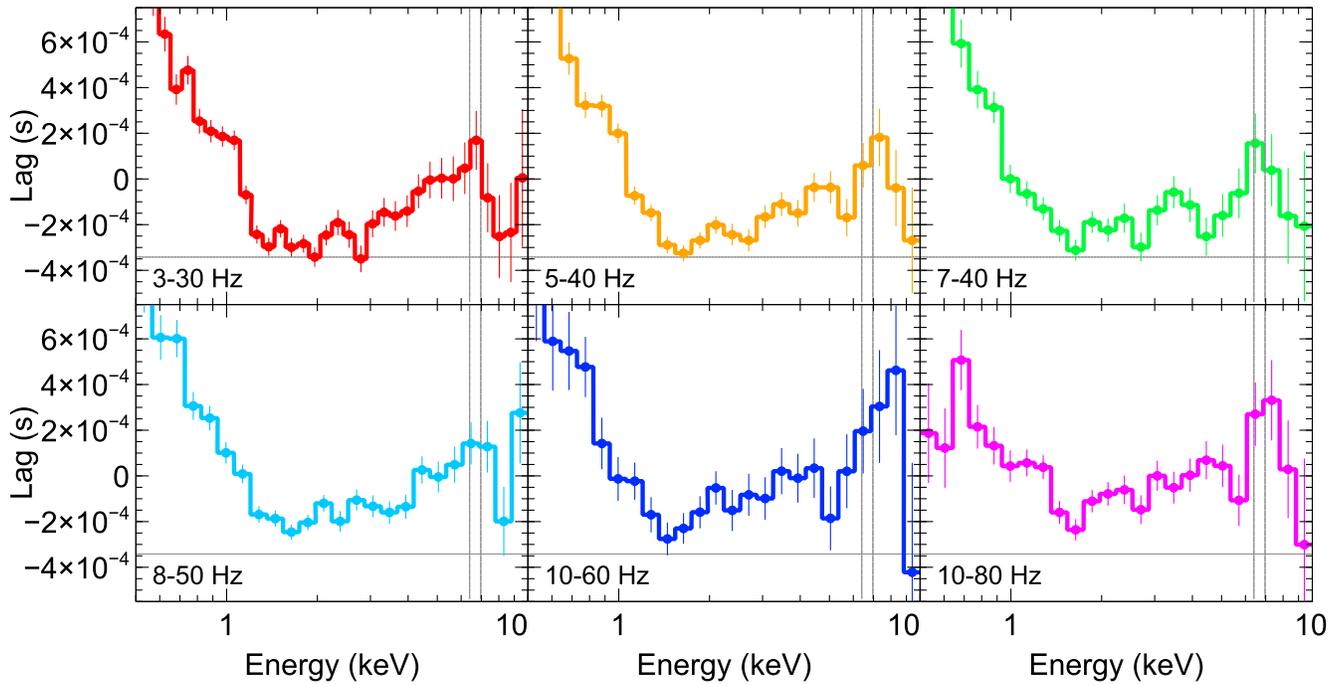}
\caption{\textbf{The high-frequency lag-energy spectra.} The evolution of the lag-energy spectra for the six observation epochs. The Fourier frequency at which the thermal lag is found increases with time (as also evident in Fig.~\ref{fig:lagfreq}). There are hints of an iron~K lag seen in these six epochs. Thin gray vertical lines indicate Fe K$\alpha$ at 6.4~keV and H-like Fe XXVI at 6.97~keV. The horizontal line indicates the minimum lag of the first epoch, as a reference by which to compare to subsequent epochs.  Error-bars indicate 1-$\sigma$ confidence intervals. }
\label{fig:lagen}
\end{figure}

To examine the high-frequency lags further, we measure the inter-band time delays by averaging over the frequencies where the soft lag is detected in Fig.~\ref{fig:lagfreq} (though extending to slightly broader frequency range for late-time epochs to maximize the signal-to-noise). The lag is measured between each small energy bin and a broad reference band, taken to be from 0.5--10~keV (with the bin-of-interest removed so that the noise is not correlated). Fig.~\ref{fig:lagen} shows the high-frequency lag-energy spectra for each of the six epochs. We see a thermal lag below 1~keV, and additionally, at higher energies, the lag peaks around the iron~K emission line at $\sim 6.4$~keV, reminiscent of the iron~K reverberation lags commonly observed in AGN systems. These iron~K lags are not all statistically significance compared to a featureless powerlaw lag (see Methods and \ref{fig:lagfit}) and are not present in all observations (\ref{fig:other_lags}), but if associated with iron~K reverberation, we find an average amplitude of $0.47\pm0.08$~ms or $14\pm3~r_{\mathrm{g}}/c$ for an assumed black hole mass of 10~$M_{\odot}$ (see Methods for a description of how the amplitude is estimated and how this translates to a light travel time delay by accounting for dilution and lags due to propagating fluctuations). The thermal reverberation lag persists at high frequencies ($\sim 10-100$~Hz) until the source transitions to the soft state and the rms variability of the source decreases (\ref{fig:other_lags}).

Previous results on the hard state of GX~339-4 observed with XMM-Newton revealed thermal lags \cite{uttley11,demarco15}, and a tentative iron~K lag\cite{demarco17} that are more than an order of magnitude larger than the reverberation lags reported here. This has been interpreted as a large truncation radius of the disc of $\sim 100 r_{\mathrm{g}}$, which is at odds with other estimates of the inner radius from spectral fitting of the gravitationally redshifted broad iron line that suggest a truncation radius of $\sim 2 r_{\mathrm{g}}$ and a coronal height of $<10 r_{\mathrm{g}}$\cite{garcia15,parker15}. NICER, with its large effective area, minimal pile-up and good spectral resolution, has revealed a consistent picture between spectral and time lag results for MAXI~J1820+070, which point to a compact corona and small truncation radius. At frequencies of less than a few Hz, the time lags in MAXI~J1820+070 are very similar in shape and amplitude to those of GX~339-4, and so we suggest it is possible that similarly short timescale iron~K reverberation would be seen in GX~339-4, if we could probe high enough frequencies to overcome the dominating continuum lag.

\begin{figure}
\centering
\includegraphics[width=\linewidth]{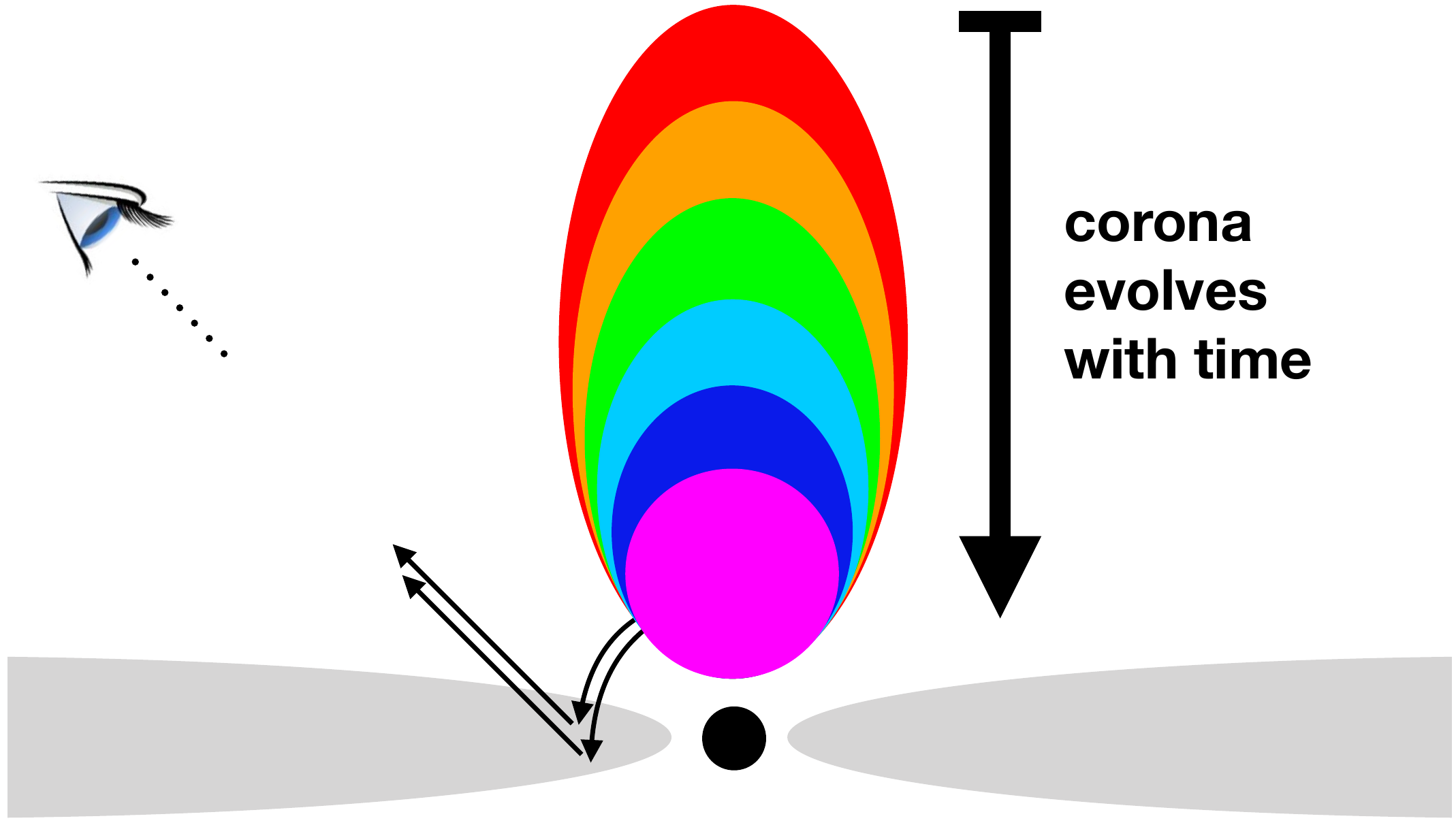}
\caption{\textbf{Schematic of the proposed geometry.} Schematic of the proposed geometry, evolving from a vertically extended corona at early times, to a more compact corona at late times. The constant shape of the broad iron line is due to a static core of the corona at small radii that is responsible for most of the flux irradiating the disc. As the corona decreases in vertical extent, the coronal variability timescale shortens, causing the shift in the thermal reverberation lag to higher frequencies. The decrease in vertical extent of the corona is also responsible for the decrease in the equivalent width of the narrow component of the iron line at 6.4~keV.}
\label{fig:schematic}
\end{figure}

The simultaneous detection of an unchanging broad iron line component (Fig.~\ref{fig:HID_spec}c) together with short reverberation lags that evolve to higher frequencies (Figs.~\ref{fig:lagfreq},\ref{fig:lagen}) suggest that the X-ray emitting region is spatially compact, and becoming more compact over time. This could be accomplished by a vertically extended corona with a compact core, which collapses down along the axis over time (see schematic in Fig.~\ref{fig:schematic}). The 3--10~keV spectra suggest a similar evolution, where, in addition to the unchanging relativistically broadened iron line, there is a second narrow 6.4~keV component that is only prominent at early times. If this narrow component is due to a vertically extended corona irradiating large radii, then as the corona collapses, the solid angle irradiating the disc at large radii decreases, thus decreasing the equivalent width of the narrow component. The fact that the thermal reverberation lags remain throughout all epochs and that the spectral shape of the broad iron line component is constant over time suggests that there is little or no evolution in the truncation radius of the inner disc during the luminous hard state. These observations of a Galactic black hole in its hard state are similar to observations of local Seyfert galaxies, which show a compact X-ray corona and a disc that extends to very small radii. 
NICER continues to take near daily observations of MAXI~J1820+070 and other Galactic black hole transients, thus providing a new tool for understanding accretion physics near the black hole event horizon.

\bibliography{sample}

\section*{Acknowledgements}
EK thanks Geoff Ryan and Peter Teuben for helpful discussions on ways to speed up her Python code and Javier Garcia and Douglas Buisson for discussions on NuSTAR observations of MAXI~J1820+070. EK acknowledges support from the Hubble Fellowship Program and the University of Maryland Joint Space Science Institute and the Neil Gehrels Endowment in Astrophysics through the Neil Gehrels Prize Postdoctoral Fellowship. Support for Program number HST-HF2-51360.001-A was provided by NASA through a Hubble Fellowship grant from the Space Telescope Science Institute, which is operated by the Association of Universities for Research in Astronomy, Incorporated, under NASA contract NAS5-26555. JFS has been supported by NASA Einstein Fellowship grant PF5-160144. EMC gratefully acknowledges NSF CAREER award AST-1351222. DA acknowledges support from the Royal Society. This work was supported by NASA through the NICER  mission and the Astrophysics Explorers Program, and made use of data and software provided by the High Energy Astrophysics Science Archive Research Center (HEASARC).

\section*{Author contributions statement}
EK led timing analysis and interpretation of results. JS produced the HID and contributed to interpretation of results. ACF performed spectral modelling and contributed to interpretation of results. EMC and PU performed cross-checks of analysis software and contributed to interpretation of results. RR contributed to background modeling and interpretation of results. KG and ZA scheduled the NICER observations and contributed to data reduction.  DA, JH, SE, TE, JN, ALS contributed to interpretation of results.

\section*{Code availability}
The model fitting of spectra and lag-energy spectra was completed with XSPEC, which is available at the HEASARC website.  The timing analysis were made with currently private Python code, however community efforts (by members of our team and others) are currently being made to aggregate Python timing analysis codes into one open source package, called stingray.  More information at github.com/StingraySoftware/stingray.  All figures were made in Veusz, the Python-based scientific plotting package, developed by Jeremy Sanders and available at veusz.github.io.

\section*{Author Information} 
Reprints and permissions information is available at www.nature.com/reprints.The authors declare no competing financial interests. Correspondence and requests for materials should be addressed to EK (ekara@astro.umd.edu).

\section*{Methods}

\subsection*{Data Reduction}

The data were processed using NICER data-analysis software (DAS) version {\sc 2018-03-01\_V003}. The data were cleaned using standard calibration with {\sc nicercal} and screening with {\sc nimaketime}. To filter out high background regions, we made a cut on the magnetic cut-off rigidity with {\sc COR\_SAX}~$> 4$. We selected events that were not flagged as ``overshoots” or ``undershoots” (EVENT FLAGS=bxxxx00) and events that were detected outside the SAA. We also omit forced triggers. We required pointing directions at least 30 degrees above the Earth limb and 40 degrees above the bright Earth limb. The cleaned events, produced with {\sc nicermergeclean}, use standard ``trumpet'' filtering to eliminate additional known background events. The cleaned events were barycenter corrected. For the spectra, we estimated in-band background from the 13--15~keV and trumpet-rejected countrates, and used this to select the appropriate background model from observations of a blank field.  To reduce the strong localized residuals that result from calibration uncertainties, the spectra of MAXI~J1820+070 were corrected using residuals from fits to the featureless power-law spectra of the Crab Nebula\cite{ludlam18}. This accounts for most of the calibration uncertainties, though NICER calibration work is ongoing. We emphasize that small uncertainties in the instrument response do not affect the time lag analysis, as the time lags are a ratio of the Imaginary and Real parts of the cross spectrum, and thus, the instrument response is divided out\cite{mastroserio18}. For the timing analysis, we bin the cleaned events in time and energy to produce uninterrupted light curve segments of 10~s duration and 0.001~s time bins in multiple energy bands.  

\subsection*{Lags in the remaining observations}

NICER continues to monitor MAXI~J1820+070 with a near daily cadence, and so detailed follow-up papers will study the lags of the entire state transition from outburst back to quiescence, but here, we briefly discuss the results from the remaining observations taken thus far.

Beyond our six epochs of interest, we analyzed all of the other observations taken between our first and last epoch (e.g. spanning from MJD~58198 to MJD~58250). We produced lag-frequency spectra between 0.5-1~keV and 1-10~keV (similar to the analysis shown for our six epochs in Fig.~\ref{fig:lagfreq}), and find clear evidence for high-frequency soft lags in all observations. \ref{fig:other_lags}-a shows the frequency range where soft lags are found. The overall trend is towards higher frequencies over time.

All high-frequency lag-energy spectra of observations between Epochs 1 and 6 show clear thermal lags, though iron~K lags are not found in all cases. 
In the observations where there are not indications of an iron K lag, it is either because of low signal-to-noise or because it appears that the hard lag continues on to the highest energies. The consistent detection of a thermal reverberation lag (where the signal-to-noise is highest) suggests that the lack of iron~K reverberation in some observations is due not a change in the reflection, but rather due to something in the continuum (or simply due to low signal-to-noise). This is consistent with our overall interpretation that the corona is driving the evolution of the source.

We analyzed the earliest observations taken as the source was rising to peak (ObsIDs: 01-05. from MJD~58189 to MJD~58193, and see dark red hashed region in the Hardness-Intensity Diagram in \ref{fig:other_lags}-b). The 0.5--1~keV vs. 1--10~keV lag-frequency spectrum shows no significant high-frequency soft lag (see \ref{fig:other_lags}-c1 for comparison to the lags in Epoch~1 near the peak luminosity).  However, examining the lag-energy spectrum in the same frequency range as Epoch~1 reveals a soft thermal lag and a dominating continuum hard lag (\ref{fig:other_lags}-c2). Despite showing a clear iron~K emission line (see inset of \ref{fig:other_lags}-c1), there is no evidence for iron~K reverberation perhaps because it is `hidden' in the strong continuum hard lag. This result is perhaps consistent with our proposed picture in which the corona is highly extended at early times, and thus dominates the lags, even up to high frequencies. 

High-frequency soft lags above 10~Hz remain throughout all of the hard state observations, even as the luminosity drops by a factor of 4. Then, at MJD~58290, MAXI~J1820+070 began a rapid transition to the soft state. It is well known in many Galactic black hole transients, that the rms variability greatly decreases as the source transitions to the soft state. We are able to measure high-frequency power above the Poisson noise limit up to MJD=58304.9 (by which time the spectral hardness has decreased from 0.29 to 0.17). Even as the source begins to transition from hard to soft state, we continue to observe soft lags at very high frequencies (see \ref{fig:other_lags}-d1 for a comparison of the lag-frequency spectrum from ObsIDs 94-96 taken from MJD~58302 MJD~58304.9 and the lag-frequency spectrum of Epoch~6 in the hard state). As the source transitions to the soft state and the rms variability decreases, the quality of the lag-energy spectra decreases, but they are consistent with the hard state lag-energy spectra (see \ref{fig:other_lags}-d2).

\subsection*{Significance tests in the lag-frequency spectra (Fig.~\ref{fig:lagfreq})}

To measure the significance of the reversal of the sign in the lag-frequency spectra shown in Fig.~\ref{fig:lagfreq}, we fit the 1--100~Hz lag-frequency spectra of all 6 epochs in {\sc xspec}. We compared two simple models: a null hypothesis with a powerlaw lag that decays to zero lag at high frequencies, and a model with a powerlaw lag plus an additional negative Gaussian to fit the high-frequency soft lag.  Comparing the change in $\chi^2$ per degree of freedom, we find the model with the additional negative Gaussian was preferred in all epochs at $4.5\sigma$ confidence or greater. 

The frequency at which the lag switches from positive to negative increases by roughly an order of magnitude. Based off the binned lag-frequency spectra in Fig.~\ref{fig:lagfreq}, the turnover frequencies for our six epochs of interest are: 2.8~Hz, 5.6~Hz, 7.9~Hz, 11.1~Hz, 15.7~Hz, 22.0 Hz. The frequency ranges from all epochs do overlap, though we strongly disfavor a solution where the frequency range of the soft lag is constant over time.  We test this by fitting all six lag-frequency spectra simultaneously with the powerlaw plus negative Gaussian model. Our null hypothesis is that the mean frequency, width and amplitude of the Gaussian is constant over all epochs, and compare this to a model where the Gaussian parameters are allowed to vary. The variable Gaussian model results in an improved fit of $\Delta \chi^{2}=150$ for 15 fewer degrees of freedom. Thus, the soft lag is increasing in frequency at $>8\sigma$ confidence. 

\subsection*{Significance tests and amplitudes in the lag-energy spectra (Fig.~\ref{fig:lagen})}

NICER allows us to probe higher frequency time lags better than ever before possible, revealing the evolution of the thermal lag to higher frequencies and potential structure at the iron~K band. Despite this advancement, we cannot be sure that we are probing frequencies beyond which the hard continuum lag no longer contributes to the lag-energy spectrum. This complicates measurements of the significance and amplitude of the iron~K lag. In this and the following section, we employ two methods: first assuming that the hard continuum lag is not contributing to the lags (i.e. the assuming that the powerlaw continuum responds simultaneously over all bands; referred to as Case A), and then assuming that the hard lag is still present (Case B). 

We fit the 0.5--10~keV lag-energy spectrum (Fig.~\ref{fig:lagen}) of all six epochs in {\sc xspec}. We fit the lag-energy spectrum with a null continuum model and compare it to the continuum plus lines model. In our case, the null continuum model is a powerlaw hard lag and a thermal lag (in {\sc XSPEC} syntax: {\sc model powerlaw+diskbb}). In Case A, the temperature and normalizations are left free to vary, and the powerlaw index is fixed at zero. In case B, the powerlaw index is also left free to vary, similar to the null hypothesis in De Marco et al., 2017\cite{demarco17}.  We then compare these to the null continuum plus a broad iron~K component and a broad iron~L component. We use the relativistically broadened iron line as prescribed in the {\sc laor} model.  For simplicity, we fix the inclination to $45 \deg$, emissivity profile to $r^{-3}$, inner disc radius to the minimum value $1.235 r_{\mathrm{g}}$, and outer disc radius to $1000 r_{\mathrm{g}}$. The normalization and line energy are the only free parameters of the model. The final significance of the results are not sensitive to the parameters of the {\sc laor} model, changing by $< 10\%$ when the inclination is left free to vary. 

\ref{fig:lagfit} demonstrates the results of lag fitting for the first observation (ObsID: 1200120106) for Case A and B. Details on the fit parameters for all 6 epochs can be found in \ref{tab:lagfit} and \ref{tab:lagfitB}. Comparing the change in $\chi^2$ per degree of freedom, we find that the addition of the iron lags in Case A is significant ($3.5\sigma$ or greater). In all cases, however, distinguishing between a pure powerlaw hard lag and a hard lag with an iron~L and iron~K lag is less clear, and the significance ranges from ($1-6.5\sigma$). The significance of the iron~L line alone is stronger than iron~K alone, as the signal-to-noise at 1~keV is much higher than in the iron~K band. These quoted significances are for the individual epochs of interest and do not account for the total number of trials from all NICER observations of MAXI~J1820+070. While we do observe peaks in the lag-energy spectra at the energies where we expect iron~L and iron~K reverberation, the features detected here are not formally significant.

Assuming that the lag features at $\sim 7$~keV are associated with iron~K reverberation lags, we measure the lag amplitude as the difference between the peak of the iron~K lag and the powerlaw normalization in that band. The Case A continuum always result in the largest lag amplitude as any hard lag contribution decreases the inferred amplitude of the lag. In the main text, we quote the Case~A lag amplitude as a 'conservative' measure of the iron~K lag. For completeness and for comparison to previous results\cite{demarco15,demarco17}, we also measure the amplitude of the thermal lag as the difference between the powerlaw normalization and the maximum lag below 1~keV. The thermal lags in MAXI~J1820+070 are both smaller in amplitude and appear at higher frequencies than in previous observations of black hole binaries with XMM-Newton\cite{uttley11,demarco15,demarco17}. See \ref{tab:lagfit} and \ref{tab:lagfitB} for the inferred thermal lag and iron~K lag from these methods. The average iron~K lag amplitude using the Case A continuum model is $0.47\pm0.08$~ms or $14\pm3$~$r_{\mathrm{g}}/c$ for a 10~$M_{\odot}$ black hole.

\subsection*{Converting the lag into a light travel distance}

Iron K lags are not a direct measure of the light travel time between the corona and the accretion disc. Effects such as the geometry of the system, inclination to the observer and relativistic Shapiro delay all play a role in the interpretation of the measured lags as physical distances. In this section, we discuss the two major contributors that are not directly related to light-travel time, but have competing effects on the interpretation of the observed lag amplitude as a light travel distance. Those are the effect of dilution (e.g., the fact that the lag is measured between energy bands, all of which contain varying contributions from broadband spectral components), and the effect of propagating fluctuations that contribute to the lags on a range of timescales. 

Dilution is caused by the presence of emission from both the driving continuum and reflection in each bin-of-interest. Its main effect is to reduce the measured amplitude of the lags by a factor of $R/(1+R)$, where $R$ is the relative amplitude of the variable, reflected flux to the variable, continuum flux in the bin-of-interest.  In AGN systems, the dilution factor increases the inferred light travel time by a factor of a few\cite{uttley14}, though in stellar-mass black holes, where the disc is typically more highly ionised and reflection fractions are lower, the effect could be significantly higher. Dilution should be treated as part of a full frequency-dependent spectral-timing model, which is beyond the scope of this work. 

Propagating fluctuations in the disc  modulate the hard X-ray emission produced through inverse Compton upscattering of thermal disc photons. These propagation lags could contribute up to the highest frequencies (e.g., up to the orbital frequency at the ISCO), and could be contributing to the interpretation of the reverberation lag at high frequencies. This effect is being explored in future works (Uttley \& Malzac, in prep. and Mahmoud, Done \& De Marco, in prep.) The effect of including propagation lags in the high-frequency lag-energy spectrum, is that the inferred light travel distance from the corona to the disc decreases by a factor of a few. 

For now, we demonstrate these effects on the time lags from the first epoch using the same technique as previous studies of the soft thermal lags\cite{demarco17}. We fit the lag-energy spectra with the same powerlaw, blackbody, 2 {\sc laor} model (as in the previous section), allowing for a non-zero powerlaw index for the continuum lag component (i.e. allowing some contribution from the continuum lag; Case~B). we measure the time delay between the continuum model and the measured lags between 6--7~keV to be $0.36\pm0.13$~ms. We then estimate the effects of dilution based on the results of spectral fitting to the time-integrated energy spectrum (Fabian et al., in preparation). The dilution factor is taken to be the ratio of the reflection component flux to powerlaw flux in the 6--7~keV band (i.e. the band over which we measured the time lag). We find $R_{6-7~\mathrm{keV}}=0.44$, which means that the intrinsic lags are reduced by a factor of $R_{6-7~\mathrm{keV}}/(1+R_{6-7~\mathrm{keV}})=0.3$. This suggests that the intrinsic lags are $\sim 1$~ms or $\sim 30~r_{\mathrm{g}}/c$ for a 10~$M_{\odot}$ black hole, though there are known caveats. Measuring the iron~K lag with respect to a continuum model implicitly assumes that there is no reflection in the reference band (an unlikely scenario), and thus the dilution factor described above and in previous works are likely lower bounds.

\subsection*{Lags in other wavebands}
X-ray time lags in both AGN and stellar-mass black holes provide information on the accretion flow at the smallest scales, closest to the black hole. Multi-wavelength time lags between X-ray, UV and optical have revealed much longer time lags that allow us to probe the accretion flow at larger scales in both AGN\cite{fausnaugh16,edelson17} and in Galactic black hole transients\cite{gandhi16,vincentelli18}. In Galactic BHBs, multiwavelength time lag analysis probes emitting regions at thousands of gravitational radii (either from reprocessing off the outer optically-emitting disc or from the IR/optical emitting part of the jet.  Joint NICER and optical monitoring campaigns of MAXI~J1828+070 are ongoing, and will be presented in future papers (Townsend et al., in prep. and Uttley et al., in prep.).

\newpage

\renewcommand\thefigure{Extended Data Figure \arabic{figure}}  
\renewcommand{\figurename}{}
\renewcommand\thetable{Extended Data Table \arabic{table}}  
\renewcommand{\tablename}{}
\setcounter{figure}{0}

\begin{figure}
\centering
\includegraphics[width=\linewidth]{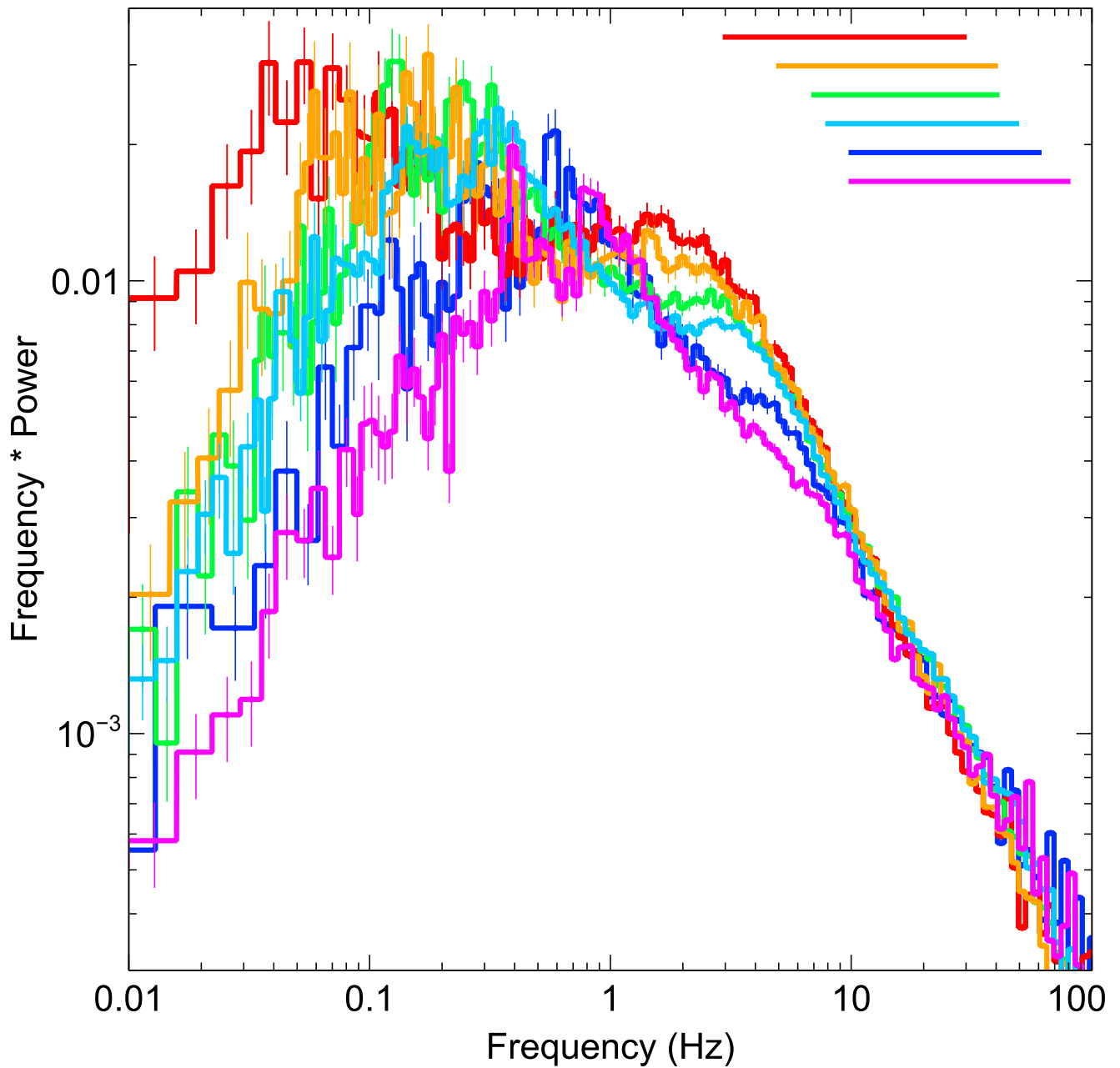}
\caption{\textbf{The power spectra evolution.} The 0.3--10~keV Poisson noise-subtracted power spectra (in units of (rms/mean)$^{2}$ ; $1-\sigma$ errors) for the six epochs of interest (same color scheme throughout). The solid lines on the top-right portion of the figure indicate the frequencies used in the lag-energy analysis (Fig.~\ref{fig:lagen}).}
\label{fig:psd}
\end{figure}

\begin{figure}
\centering
\includegraphics[width=\linewidth]{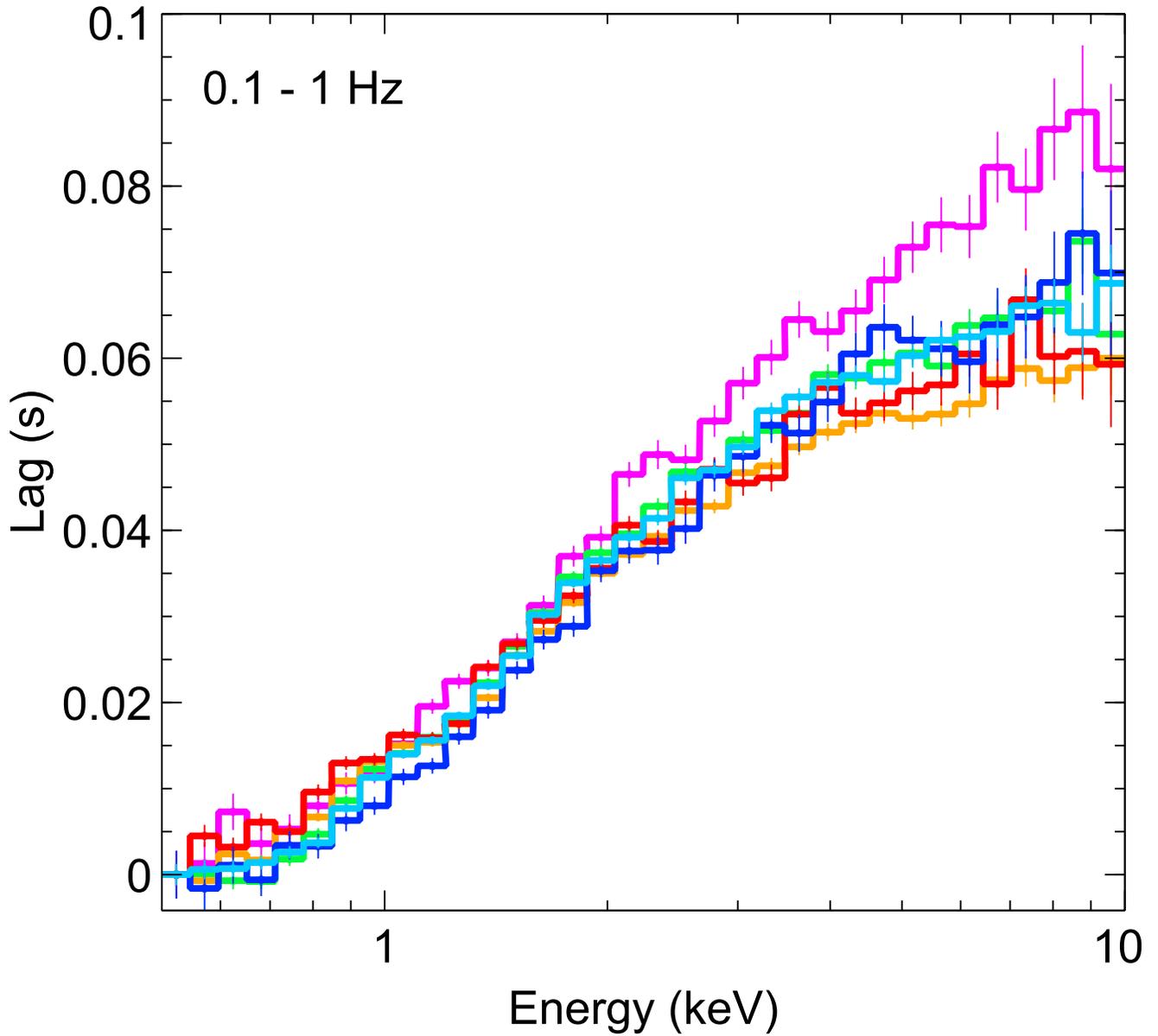}
\caption{\textbf{The low-frequency lag-energy spectra.} The low-frequency (0.1--1~Hz) lag-energy spectra for the six epochs. The lags have been shifted such that the lowest energy lag starts at zero. No thermal lag is seen at low frequencies. Error-bars indicate 1-$\sigma$ confidence intervals.}
\label{fig:low_freq_lagen}
\end{figure}

\begin{figure}
\centering
\begin{subfigure}{.5\textwidth}
\centering
\includegraphics[width=\linewidth]{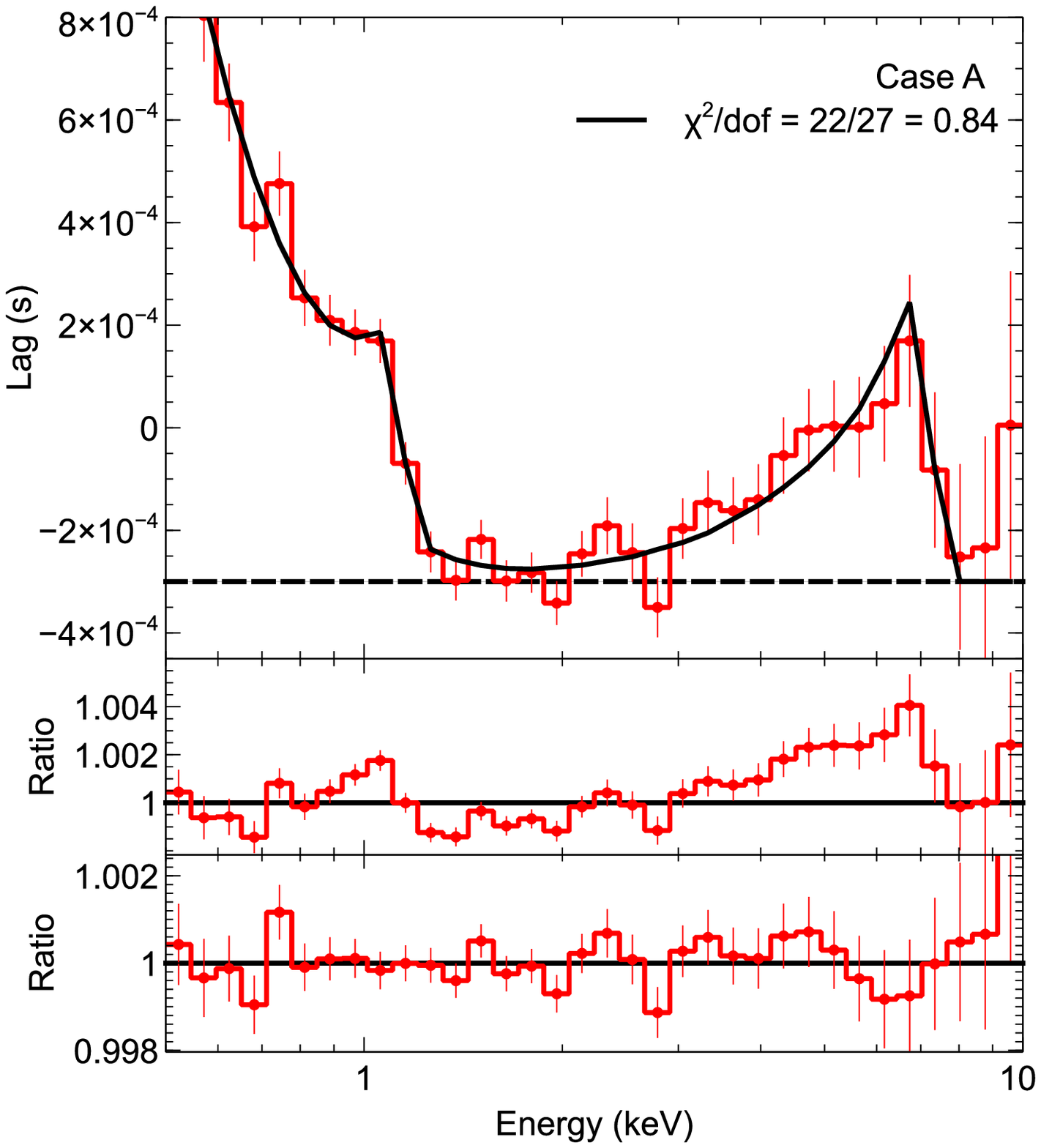}
\end{subfigure}%
\begin{subfigure}{.5\textwidth}
  \centering
  \includegraphics[width=\linewidth]{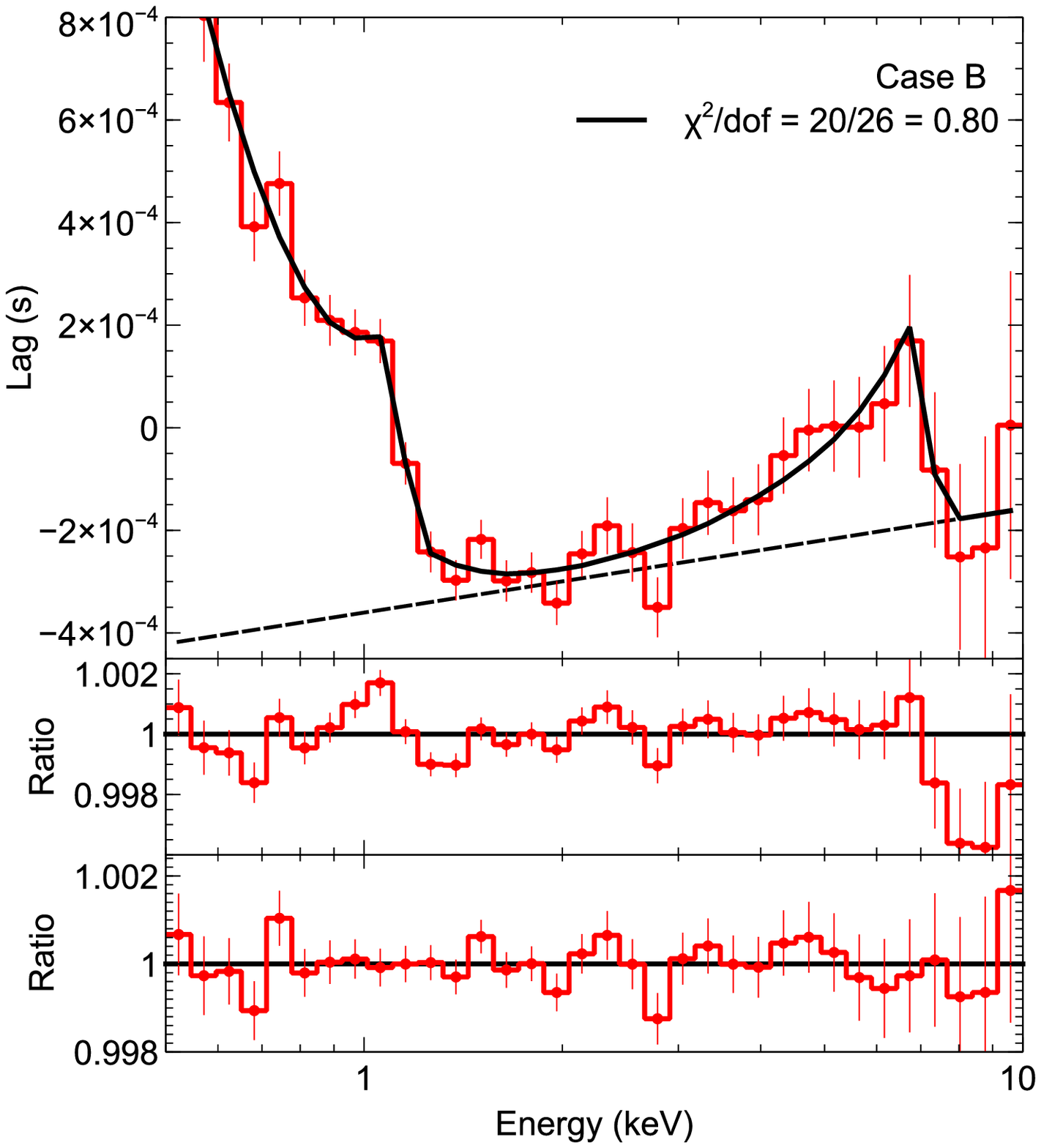}
\end{subfigure}
\caption{\textbf{Modelling the lag-energy spectra.} Top panels: The best fit Case A and Case B models fit to Epoch 1 (obsid: 1200120106), demonstrating how the significance and amplitude of the iron~K lag are determined.  Middle Panels: The ratio of the data to the null hypothesis ({\sc powerlaw+diskbb}). The Case A null hypothesis is assuming a contintuum lag powerlaw index of zero, while Case B allows for a non-zero continuum lag. Bottom panels: The ratio of the data to the best fit model ({\sc powerlaw+diskbb+laor+laor}), again where Case B allows for a non-zero powerlaw continuum lag. See text and \ref{tab:lagfit} for details. Error-bars indicate 1-$\sigma$ confidence intervals.}
\label{fig:lagfit}
\end{figure}

\begin{figure}
\centering
\includegraphics[width=\linewidth]{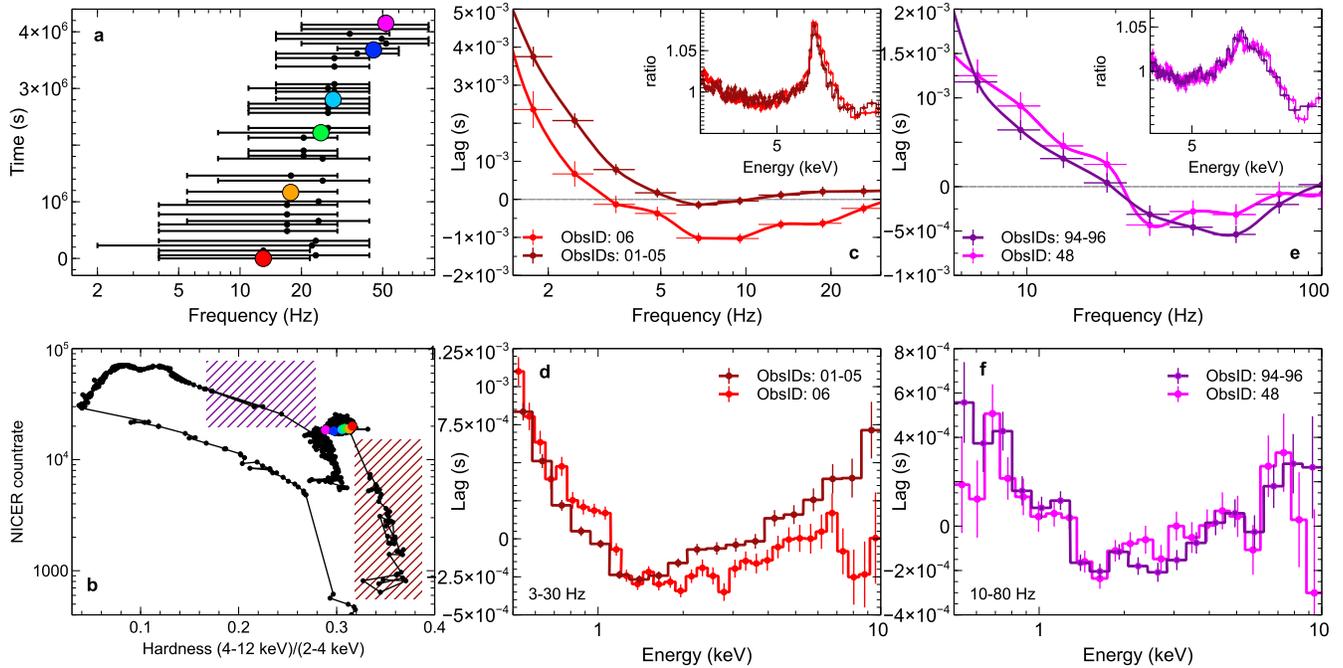}
\caption{\textbf{Lags from other observations.} (a) The frequency range of the high-frequency soft lags (lags between 0.5--1~keV and 1--10~keV) for all observations between Epoch 1 and 6. The general trend is that the soft lags increase to higher frequencies over time. The colored dots show the frequency ranges for our 6 epochs of interest (b) the Hardness-Intensity Diagram, defined as the total 0.2--12 keV count rate vs. the ratio of hard (4--12 keV) / soft (2--4 keV) count rates (same as Fig.~\ref{fig:HID_spec}-c) for all available data up to MJD~58344. This extended Hardness-Intensity Diagram shows the recent transition to the soft state. In the right two panels, we show the lags from the earliest observations from the beginning of the outburst (dark red hashed region) and from the latest times where we can measure high-frequency time lags, at the beginning of the transition to the soft state (purple hashed region). (c1) Comparison of the lag-frequency spectrum of the first epoch (ObsID: 06) and the five co-added ObsIDs that preceded it (MJD~58189 to MJD~58193). The inset shows a comparison of the ratio of the energy spectra in these epochs to a powerlaw fit from 3--10~keV. (c2) The corresponding lag-energy spectra for the 3--30~Hz range, where iron~K lags were seen in Epoch~1. The earlier observations (01-05) show a dominating hard lag at high energies, and no evidence for iron~K lags. (d1/d2) Same as (c1/c2), but comparing the lag-frequency spectra and lag-energy spectra of Epoch 6 to later observations as the source begins to transition to the soft state. Error-bars indicate 1-$\sigma$ confidence intervals.}
\label{fig:other_lags}
\end{figure}

\newpage

\begin{table}
\centering
\begin{tabular}{|c|c|c|c|c|}
\hline
Epoch & Date & ObsID & Exposure (s) & 0.2-12 keV Count rate (cts/s) \\
\hline
1 & 2018-03-21 & 1200120106 & 5438 & 20568 \\
\hline
2 & 2018-04-04 & 1200120120 & 6487 & 19015 \\
\hline
3 & 2018-04-16 & 1200120130 & 10619 & 18931 \\
\hline
4 &2018-04-21 & 1200120134 & 6964 & 18487 \\
& 2018-04-23 & 1200120135 & 3692 & 18731 \\
\hline
5 & 2018-05-02 & 1200120142 & 5512 & 17983 \\
\hline
6 & 2018-05-08 & 1200120148 & 4260 & 18403 \\
\hline
\end{tabular}
\caption{\textbf{Overview of the observations used in this analysis.} The count rate is for 52 active detectors.}
\label{tab:obs} 
\end{table}

\begin{table}
\footnotesize
\centering
\begin{tabular}{|c|c|c|c|c|c|c|}
\hline
\multicolumn{7}{|c|}{Case A}\\
 \hline
Epoch & 1 & 2 & 3 & 4 & 5 & 6 \\
\hline
$A_{\mathrm{po}}$ ($\times 10^{-4}$ s) & $-3.0 \pm 0.3$ & $-2.7 \pm 0.3$ & $-2.4\pm 0.2$ & $-2.1 \pm 0.3$ & $-2.0 \pm 0.6$ & $-1.5 \pm 0.4$ \\
$T_{\mathrm{diskbb}}$ (eV) & $20 \pm 4$  & $20 \pm 4$ & $21^{+4}_{-6}$ & $22 \pm 5$ & $19 \pm 5$ & $23^{+14}_{-8}$\\
$E_{\mathrm{laor1}}$ (keV) & $6.5^{+0.2}_{-0.4}$ & $7.3^{+0.4}_{-0.6}$ & $6.8^{+1}_{-0.4}$ & $6.7^{+0.5}_{-0.3} $ & $7.8^{+0.9}_{-0.6}$ & $7.2^{+0.5}_{-0.7}$\\
$E_{\mathrm{laor2}}$ (keV) & $1.02 \pm 0.02$ & $0.98^{+0.02}_{-0.04}$ & $0.85^{+0.50}_{-0.09}$ & $1.01^{+0.07}_{-0.04}$& $0.77^{+0.13}_{-0.08}$ & $1.18^{+0.08}_{-0.06}$\\
\hline
$\Delta\chi^2$/d.o.f. & 105/4 & 69/4 & 20/4 & 53/4& 19/4 & 27/4\\
\hline
Fe~K lag amplitude ($\times 10^{-4}$~s)& $4.7\pm 1.3$ & $4.5\pm 1.3$ & $4.0\pm 2.4$& $3.6\pm 0.9$ & $6.6\pm2.9$ & $4.8\pm 1.8$ \\ 
Thermal lag amplitude ($\times 10^{-4}$~s)& $14.0\pm 1.0$ & $14.0\pm 1.1$ & $17.1\pm 1.6$& $10.5\pm 1.3$ & $10.7\pm2.9$ & $6.5\pm 1.4$ \\
\hline
\end{tabular}
\caption{\textbf{Fit parameters of the Case A model.} Results of the lag fitting to measure the iron~K lag amplitude and thermal lag amplitude. $A_{\mathrm{po}}$ is the lag amplitude of the continuum component, modeled as a powerlaw with index fixed at zero (Case A). $T_{\mathrm{diskbb}}$ is the temperature of the inner disk for a multi-color disk blackbody component. $E_{\mathrm{laor1}}$ and $E_{\mathrm{laor2}}$ are the energies of the {\sc laor} components. The iron~K lag and thermal lag amplitude are the difference between the peak of the iron~K and thermal lag components and $A_{\mathrm{po}}$. Error bars represent 90\% confidence intervals.}
\label{tab:lagfit} 
\end{table}

\begin{table}
\footnotesize
\centering
\begin{tabular}{|c|c|c|c|c|c|c|}
\hline
\multicolumn{7}{|c|}{Case B}\\
 \hline
Epoch & 1 & 2 & 3 & 4 & 5 & 6 \\
\hline
$A_{\mathrm{po}}$ ($\times 10^{-4}$ s) & $-3.6^{+0.7}_{-1.0}$ & $-3.7 \pm 1$ & $-3.3 \pm 0.2$ & $-3.1 \pm 0.9$ & $-2.6^{+1.4}_{-2.2}$ & $-2.5 \pm 1.0$ \\
$\Gamma_{\mathrm{po}}$ ($\times 10^{-3}$) & $0.89 \pm 0.11$ & $1.3^{+1.2}_{-1.3}$ & $0.5\pm 1.5$ & $1.3^{+1.0}_{-1.2}$ & $1.2^{+1.9}_{-2.2}$ & $1.3 \pm 1.5$ \\
$T_{\mathrm{diskbb}}$ (eV) & $23^{+7}_{-5}$  & $25^{+7}_{-6}$ & $22^{+7}_{-9}$ & $26^{+8}_{-6}$ & $24 \pm 12$ & $28^{+18}_{-11}$\\
$E_{\mathrm{laor1}}$ (keV) & $6.4^{+0.3}_{-0.4}$ & $7.2^{+0.8}_{-1.2}$ & $6.8^{+1.1}_{-0.6}$ & $6.8^{+0.7}_{-0.8} $ & $7.9^{+0.07}_{-1.9}$ & $7.1^{+0.9}_{-1.0}$\\
$E_{\mathrm{laor2}}$ (keV) & $1.02 \pm 0.02$ & $0.98^{+0.02}_{-0.04}$ & $0.84^{+3.0}_{-2.3}$ & $1.02^{+0.08}_{-0.05}$& $0.76^{+9.0}_{-0.2}$ & $1.18^{+0.08}_{-0.06}$\\
\hline
$\Delta\chi^2$/d.o.f. & 44/4 & 19/4 & 4/4 & 10/4& 4/4 & 12/4\\
\hline
Fe~K lag amplitude ($\times 10^{-4}$~s)& $3.6 \pm 1.3$ & $2.8\pm1.3$ & $3.5\pm 1.3$& $2.1 \pm 0.9$ & $3.2\pm2.5$ & $3.1 \pm 1.8$ \\ 
Thermal lag amplitude ($\times 10^{-4}$~s)& $15.2\pm 0.9$ & $15.9\pm 1.0$ & $17.7\pm 1.6$& $12.2\pm 1.2$ & $12.1\pm2.9$ & $8.1\pm 1.3$ \\
\hline
\end{tabular}
\caption{\textbf{Fit parameters of the Case B model.} Same as \ref{tab:lagfit}, but for null continuum model Case B, where the powerlaw index $\Gamma_{\mathrm{po}}$ is free to vary.}
\label{tab:lagfitB} 
\end{table}

\end{document}